\documentclass[a4paper,twocolumn]{esapub} 
\usepackage{natbib}  
  
\title{Integral/IBIS Observations of persistent black hole spectral  
states during the Core Program}  
\author[1]{P. Laurent}  
\author[1]{P. Goldoni}  
\author[1]{A. Goldwurm}  
\author[1]{F. Lebrun}  
\affil{CEA/DSM/DAPNIA/Sap, CEA Saclay, F-91191  
Gif-Sur-Yvette, France}

\input{psfig.sty}  
\begin{document}  
  
\maketitle  
 
\keywords{Black hole candidates; Core Program; Integral/IBIS}

\begin{abstract}  
  
The Imager on Board the Integral Satellite (IBIS) is one of the two main  
telescopes of Integral, the ESA soft $\gamma$-ray mission to be launched in 2002. The Integral Core Program will be divided into two main parts, the Galactic Centre Deep Exposure program and the Galactic Plane Scan for a total amount of around $6.6 10^6$ seconds each year. In this paper, we will study the visibility of persistent galactic black holes as observed by IBIS during these two phases of the Core Program.  
We will also present what information may be derived from the IBIS observations of spectral/temporal properties (variation of the spectral index, presence/absence of a spectral break, ...) of these binary systems in different spectral states, in particular in the framework of the bulk motion Comptonization model.  
  
\end{abstract}

\section{The Integral satellite and its instruments}

\noindent Integral is a 15 keV-10 MeV $\gamma$-ray mission with concurrent  
source monitoring at X-rays (3-35 keV) and in the optical range (V, 500-  
600 nm). All instruments are coaligned and have a large FOV, covering  
simultaneously a very broad range of sources. The Integral payload consists  
of two main $\gamma$-ray instruments, the spectrometer SPI and the imager IBIS, 
and of two monitor instruments, the X-ray monitor JEM-X and the Optical  
Monitoring Camera OMC.  
  
  
\noindent The Imager on Board Integral Satellite (IBIS) provides diagnostic  
capabilities of fine imaging (12' FWHM), source identification and spectral  
sensitivity to both continuum and broad lines over a broad (15~keV--10~MeV)  
energy range. It has a continuum sensitivity of 2~10$^{-7}$~ph~cm$^{-2}$  
~s$^{-1}$ at 1~MeV for a 10$^6$ seconds observation and a spectral resolution  
better than 7~$\%$ @ 100~keV and of 6~$\%$ @ 1~MeV. The imaging capabilities of  IBIS are characterized by the coupling of its source discrimination capability  
(angular resolution 12' FWHM) with a field of view (FOV) of 9$^\circ$  
$ \times $ 9$^\circ$ fully coded and 29$^\circ$ $ \times $ 29$^\circ$ partially 
coded.  
  
\noindent The IBIS detection system is composed of two planes, an upper layer made of 16384 squared CdTe pixels (ISGRI) with higher efficiency below about 200 keV and a lower layer made of 4096 CsI scintillation bars (PICsIT) more efficient above 200 keV. A photon may interact with only one of the two layers giving rise to an ISGRI or a PICsIT event (the PICsIT event can be single or multiple).  
If it interacts with both, undergoing a Compton scattering, its energy  
and arrival direction can be reconstructed leading to the definition of a third type of event, the Compton one which again can be single or multiple depending on interaction in PICsIT.  
  
  
\noindent The spectrometer SPI will perform spectral analysis of $\gamma$  
ray point sources and extended regions with an unprecedented energy  
resolution of $\sim$ 2 keV (FWHM) at 1.3 MeV. Its large field of view  
(16$^{\circ}$ circular) and limited angular resolution ( 2$^{\circ}$ FWHM)  
are best suited for diffuse sources imaging but it retains nonetheless the  
capability of imaging point sources. It has a continuum sensitivity of  
7 $\times$ 10$^{-8}$ ph cm$^{-2}$ s$^{-1}$ at 1 MeV and a line sensitivity  
of 5$\times$ 10$^{-6}$ ph cm$^{-2}$ s$^{-1}$ at 1 MeV, both 3$\sigma$ for  
a 10$^6$ seconds observation.  
  
  
\noindent The Joint European Monitor JEM-X supplements the main Integral  
instruments and provides images with 3' angular resolution in a 4.8$^{\circ}$  
fully coded FOV in the 3-35 keV energy band. The Optical Monitoring Camera  
(OMC) will observe the prime targets of Integral main $\gamma$ ray instruments.  
Its limiting magnitude is M$_V$ $\sim$ 19.7 (3$\sigma$, 10$^3$ s). The wide  
band observing opportunity offered by Integral provides for the first time  
the possibility of simultaneous observations over 7 orders of magnitude  
in energy.  
  
\section{The Integral Core Program}  
 
The Integral observing time will be divided into two main parts, the Open Time devoted to Guest Observers selected by the Integral Time Allocation Committee, and the Guaranteed Time given to the member of the Integral Science Working Team, representing mostly laboratories which have worked on the Integral program. 
 
\begin{table} 
\label{Table1} 
\[ 
    \begin{array}{ccccc} 
    \hline 
\noalign{\smallskip} 
  ${\rm Year }$ & ${\rm Total} $ & ${\rm GPS}$ & ${\rm GCDE}$ & ${\rm P.O.}$\\ 
\noalign{\smallskip} 
\hline 
\noalign{\smallskip} 
1  & 9.32 & 4.30 & 2.30 & 2.72 \\ 
2  & 7.98 & 4.30 & 2.30 & 1.38 \\ 
3+ & 6.64 & 4.30 & 1.47 & 0.87 \\ 
\hline 
\end{array} 
\] 
\caption{Available observing time in the different parts of the Core Program in $10^6$ seconds unit.} 
\end{table} 
  
\begin{table} 
\label{Table2} 
\[ 
    \begin{array}{ccc} 
    \hline 
\noalign{\smallskip} 
  ${\rm Name }$ & ${\rm Limit Flux} $ & ${\rm Energy (keV)}$\\ 
\noalign{\smallskip} 
\hline 
\noalign{\smallskip} 
 ${\rm GRS 1915+105  }$ & ${\rm 300 mCrab }$ & 30  \\ 
 ${\rm GROJ 1655-40  }$ & ${\rm 300 mCrab }$ & 100  \\ 
 ${\rm 1E 1740.7-2942}$ & ${\rm 200 mCrab }$ & 100  \\ 
 ${\rm Cyg X-1       }$ & ${\rm 400 mCrab }$ & 30  \\ 
 ${\rm Cyg X-3       }$ & ${\rm 400 mCrab }$ & 30  \\ 
 ${\rm GX 339-4      }$ & ${\rm 400 mCrab }$ & 30  \\ 
\hline 
\end{array} 
\] 
\caption{List of the selected Target of Opportunity for the Core Program, with the Flux limit at the given Energy above which a TOO is declared.} 
\end{table}

The Integral guaranteed time (also called the Integral Core Program) will be in turn divided into three parts, the Galactic Plane Survey, the Galactic Central Radian Deep Exposure program and some Target of Opportunity (TOO) sources (see Winkler et al., this conference, for more details). Table 1 gives the observing time available for these different parts of the Core Program during the whole mission. 
 
The Galactic Plane Survey will consist on a scanning of the galactic  
plane once a week. The pointing will follow a saw-tooth pattern, with  
an observing time on each point of 1050 s. As a given source will be detected 
in the large field of view of the major Integral instrument during three successive pointing, the total exposure for a given source is around 3000 seconds each week. 
 
The Galactic Central Radian Deep Exposure will be a deep survey of the central Galactic radian. The total observing time for one year will be $2.3 10^6$ seconds. Lastly, some TOO sources, given in Table 2, will be followed up, if their flux at a given energy rise above a certain threshold, and if there is time remaining in the Core Program to allocate these observations.  
  
\section{Black holes spectral states}   
 
The Black Hole Systems (BHS) are generally observed in two spectral  
states: one  with a high luminosity soft thermal bump and a  
low luminosity power law extending to several hundred of keV  
(hereafter called the ``soft state" see Figure 1), and the other at  
high luminosity in  high energies ($> 10$ keV) showing  
a typical thermal Comptonization component (called the ``hard state" in this paper, see Figure 2). The origin of the powerlaw in the soft state is still 
under debates, but it could result from the Comptonization of soft X-ray  
photons from the accretion disk, by the free-falling electrons onto the 
black hole (Laurent \& Titarchuk 1999, hereafter LT99). In that optics, the state transition, soft to hard, will be caused only by an increase in the Comptonizing plasma temperature. Indeed, the electrons thermal motion would be dominated in the soft state by the free-falling motion. Conversely, the reverse situation would occur in the hard state. We will now see what information we can get from these systems during the GPS and the GCDE. 
 
\begin{figure}  
\centerline{\psfig{figure=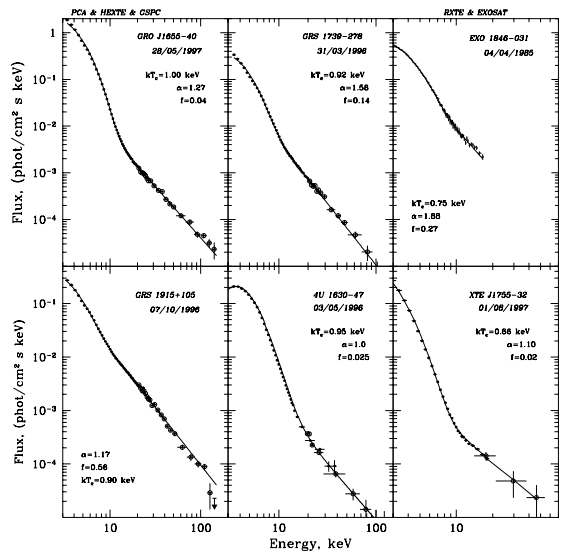,width=80mm}}  
\caption{Plot of different Black Hole Systems (GRO J1655-40, GRS
1915+105, GRS 1739-278, 4U 1630-47 XTE J1755-32, and EXO 1846-031) in the low state, as observed by RXTE and EXOSAT (Borozdin et al. 1999)}  
\end{figure}  
  
\begin{figure}  
\centerline{\psfig{figure=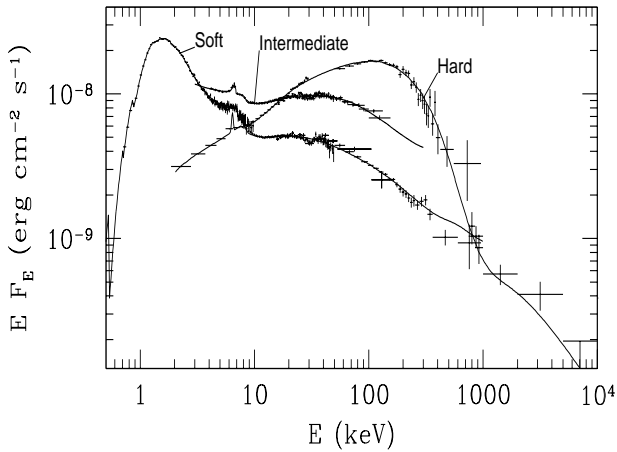,height=60mm,width=80mm}}  
\caption{The different spectral states of Cygnus X-1 (from Gierlinski et al. 1999)}  
\end{figure}  
 
 \section{Observations of persistent Black Holes during the Galactic Plane Survey}  
 
Table 3 gives the significance of detection of the main persistent Black Hole Systems during one GPS scan. It could be seen clearly on this Table that due the high level of detection we will get, the spectral state of these sources will be clearly determined. This is furthermore confirmed by the simulations we made of one GPS scan with the IBIS Mass Model (see Laurent et al., this conference, for a description).  
 
Indeed, we have modelled first the observation of a 300 mCrab Black Hole System in the soft state, by using a powerlaw input photon spectrum with an index of -2.9 between 20 and 1000 keV, which is representative of what is observed from these systems (LT99). Then, we made a simulation with a system of similar brightness, but using this time a powerlaw of photon index -0.8, representative of the thermal Comptonization bump observed in the hard state (Ebisawa et al., 1996). Figure 3 and 4 shows the result of these simulations for the ISGRI events, and we can see there that we could clearly discriminate between the different black hole spectral states within one GPS scan. Also, it is worth noting that, due to the high sensitivity of IBIS/ISGRI below 150 keV and also due to its 64 ${\mu}s$ timing resolution, the study of QPOs appearance, spectral dependence, and variability's will be easily done during each GPS scan. 
 
\begin{table} 
\label{Table3} 
\[ 
    \begin{array}{ccc} 
    \hline 
\noalign{\smallskip} 
  $$ & ${\rm detection level}  $ & \\ 
 ${\rm Name }$ & ${\rm low state} $ & ${\rm high state}$\\ 
\noalign{\smallskip} 
\hline 
\noalign{\smallskip} 
 ${\rm Cyg X-1       }$ & ${\rm $40 \sigma$ }$ & ${\rm $100 \sigma$ }$ \\ 
 ${\rm GRS 1915+105  }$ & ${\rm $10 \sigma$ }$ & ${\rm $45 \sigma$ }$   \\ 
 ${\rm 1E 1740.7-2942}$ & ${\rm $ < 3 \sigma$ }$ & ${\rm $15 \sigma$ }$ \\ 
 ${\rm GRS 1758-258  }$ & ${\rm $ < 3 \sigma$ }$ & ${\rm $10 \sigma$ }$ \\ 
 ${\rm GX 339-4      }$ & ${\rm $5 \sigma$ }$ & ${\rm $60 \sigma$ }$    \\ 
\hline 
\end{array} 
\] 
\caption{Significance level of detection of the main persistent Black Hole Systems during one GPS scan (that is for a observing time of the source of 3250 s.).} 
\end{table}

\begin{figure}  
\centerline{\psfig{figure=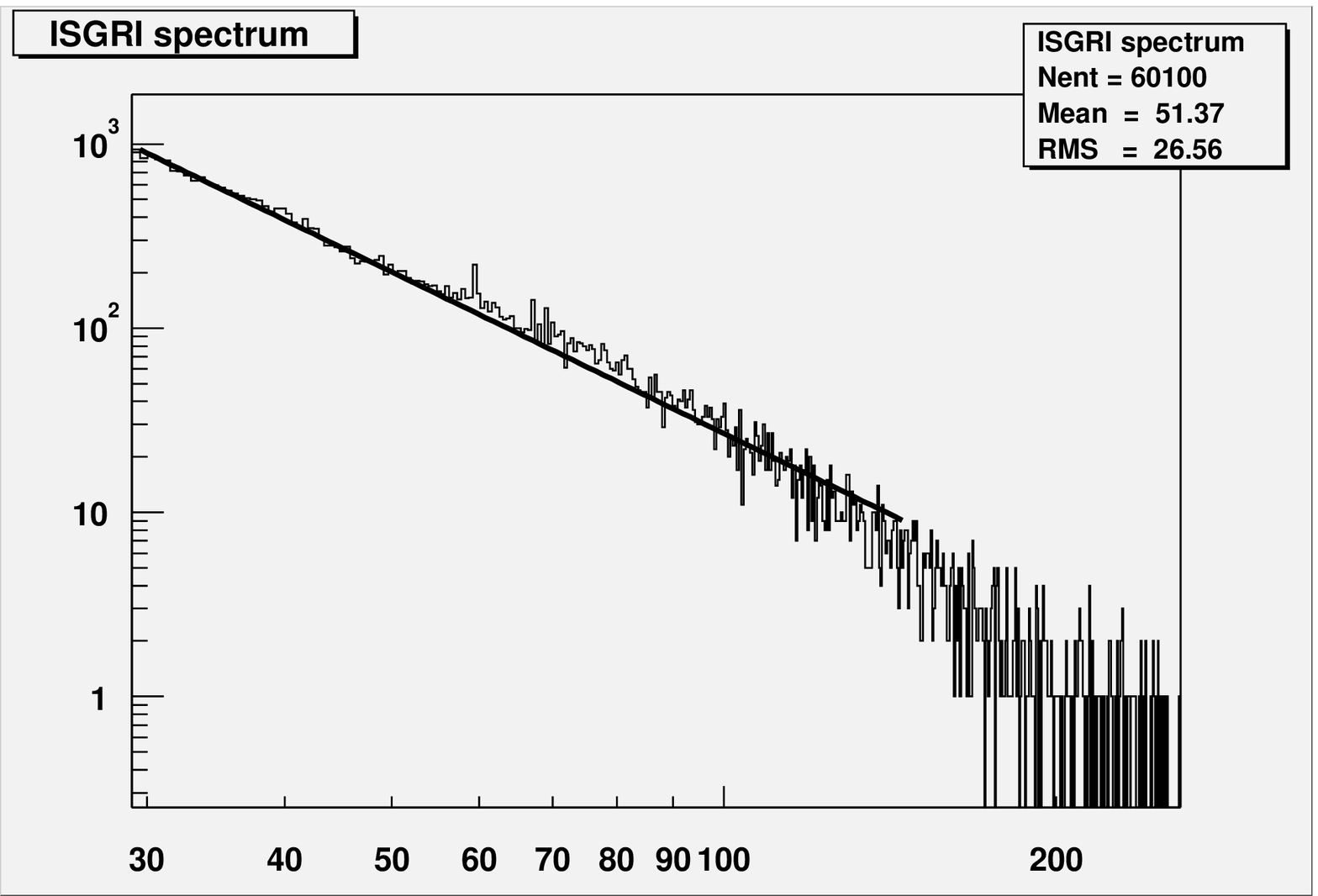,width=80mm}}  
\caption{ISGRI simulated count spectrum of a 300 mCrab Black Hole System in the soft state acquired during one GPS scan. The powerlaw fit also shown has a photon index of -2.9. The lines seen in the spectrum are background lines due to the lead and tungsten fluorescence in the telescope.}  
\end{figure}

\begin{figure}  
\centerline{\psfig{figure=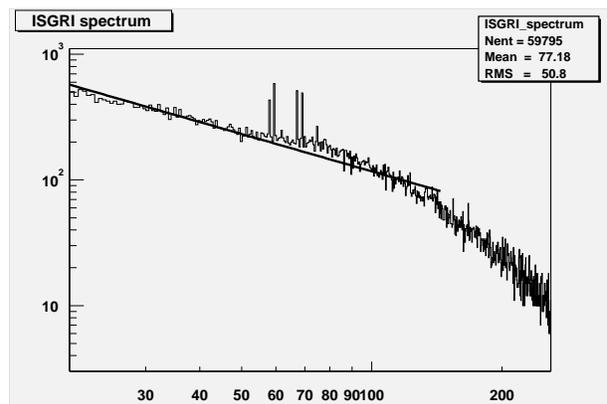,width=80mm}}  
\caption{ISGRI simulated count spectrum of a 300 mCrab Black Hole System in the hard state acquired during one GPS scan. The powerlaw fit also shown has a photon index of -0.99. The lines seen in the spectrum are background lines due to the lead and tungsten fluorescence in the telescope.}  
\end{figure}  
  
\section{Observations of persistent Black Holes during the Galactic Central Radian Deep Exposure}

During one year of GCDE, we will get with IBIS a sensitivity of the order of the ones given by Figure 5. We can clearly see there that the spectral continuum of a Black Hole System in the central radian will be clearly determined up to a few hundredth of keV. Also, we can obtain the same kind of sensitivity for GPS sources, if we sum their observations during the whole Integral mission. To do this, we shall expect however that the source spectrum do not changed much during the mission, which seems not very plausible. 
 
Knowing this sensitivity, we can try to use the IBIS high energy data in order to determine if there is a spectral break in Black Hole System spectra around 400 keV, as it is foreseen in the Bulk Motion Comptonization model. To test this, we have made a simulation of a 400 mCrab system in the soft state (powerlaw of photon index -2.9), and we show in Figure 6 the simulated count PiCsIT (single + multiple events) spectrum, taking into account a predicted background of 5000 cts/s. The observing time simulated was $2.3  10^6$ seconds, that is one year of GCDE. As it could be seen in Figure 6, the statistics in that case is clearly not enough to analyse a possible break around 400-500 keV. This feature may be detected in fact only if we consider a stronger source (around one Crab), and if we consider also that we can increase our sensitivity by using the SPI data. Such a strong source has however never been observed in the central radian, so the only study of that kind we can foresee for the moment, using the Core Program data, is the study of Cygnus X-1 using the GPS data acquired during the whole mission. {\it So, in our opinion, the study of the $> 400$ keV part of the spectrum of  hundredth of mCrab sources in the soft state should be made during the Open Time.}

\begin{figure}  
\centerline{\psfig{figure=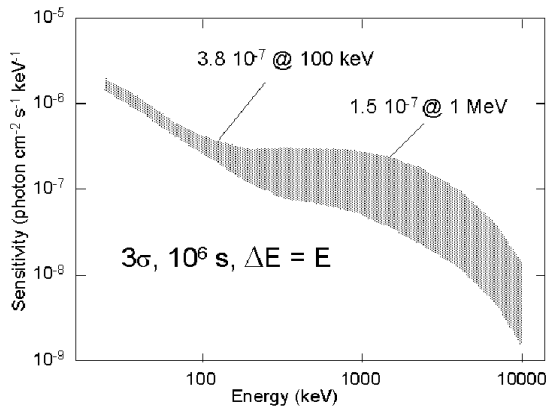,width=80mm,angle=270}}  
\caption{IBIS $3 \sigma$ sensitivity for an $10^6$ seconds observation, representative of one year of GCDE.}  
\end{figure}  
   
\begin{figure}  
\centerline{\psfig{figure=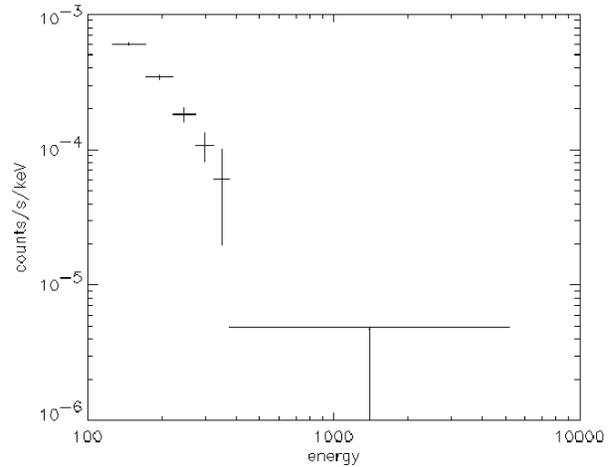,width=80mm}}  
\caption{PiCsIT spectrum of a 400 mCrab Black Hole System in the soft state, with the background included (see text).}  
\end{figure}

\end{document}